\documentclass[12pt,preprint]{aastex}

\def\au{{\rm AU}}

\begin{document}
\title{Finding Planets Around White Dwarf Remnants of Massive Stars}

\author{Andrew Gould and Mukremin Kilic}
\affil{Department of Astronomy, Ohio State University,
140 W.\ 18th Ave., Columbus, OH 43210, USA; 
gould,kilic@astronomy.ohio-state.edu}

\begin{abstract}

Planet frequency shows a strong positive correlation with host mass
from the hydrogen-burning limit to $M\sim2\,M_\odot$.  No search
has yet been conducted for planets of higher-mass hosts because
all existing techniques are insensitive to these planets.
We show that infrared observations of the white-dwarf (WD) remnants
of massive stars
$3\,M_\odot \la M\la 7\,M_\odot$ would be sensitive
to these planets for reasons that are closely connected to the
insensitivity of other methods.  We identify 49 reasonably
bright, young, massive WDs from the Palomar-Green survey and discuss
methods for detecting planets and for distinguishing between planet
and disk explanations for any excess flux observed.  The young,
bright, massive WD sample could be expanded by a factor 4--5 by
surveying the remainder of the sky for bright UV-excess objects.

\end{abstract}


\section{Introduction}

Extrasolar planets have been discovered by 4 different techniques
orbiting stars ranging from the hydrogen-burning limit to about
2.5 solar masses. As first pointed out by \citet{laws03},
the fraction of monitored stars that are found to have 
planets climbs steadily as the host mass increases from 
$0.75\,M_\odot$ to $1.5\,M_\odot$.  \citet{fischer05} found
a more significant trend in a larger sample and noted that while
this might reflect
a higher rate of planet formation around massive stars, it
could also be an indirect selection effect due to the
known correlation between planet frequency and metal abundance,
combined with the fact that their sample
shows a strong correlation between stellar mass and metallicity.
\citet{johnson07} extended this correlation to the entire observed
range of host masses $0.1\,M_\odot< M < 1.9\,M_\odot$
(considering only planet masses $m>0.8\,M_{\rm jup}$ and semimajor
axes $a<2.5\,$AU)
 and argued
that it was in fact independent of the correlation of frequency with
metallicity.  

It would be quite interesting to determine whether the correlation
of planet frequency with host mass continues to yet higher masses.
However, current techniques are extremely challenged in the regime 
$M\ga 3\,M_\odot$.
Radial-velocity (RV) searches are restricted to FGKM stars
because of the absence of usable lines in earlier-type stars.
RV searches have managed an ``end run'' around this problem
for stars of 1--3 $M_\odot$ by monitoring the {\it red giant descendants}
of these early type stars 
\citep{frink02,reffert06,johnson07,lovis07}. 
Reasonable mass estimates can be
obtained from spectroscopic gravities, although these are not
quite as accurate as the mass estimates of main-sequence (MS) stars.

Massive red giants still pose significant challenges for the RV
technique.  Their low gravities induce atmospheric instabilities,
which radically degrade RV precision relative
to what is possible for stable MS stars.  So far,
this reduced precision has limited searches to relatively
massive planets with periods of order a few hundred days, i.e.,
well inside the ``snow line'' that is believed to be the
boundary of massive-planet formation.  Hence, these searches
probably probe only migrated planets and so provide a much
smaller window than is available for less massive stars.

\begin{figure}
\plotone{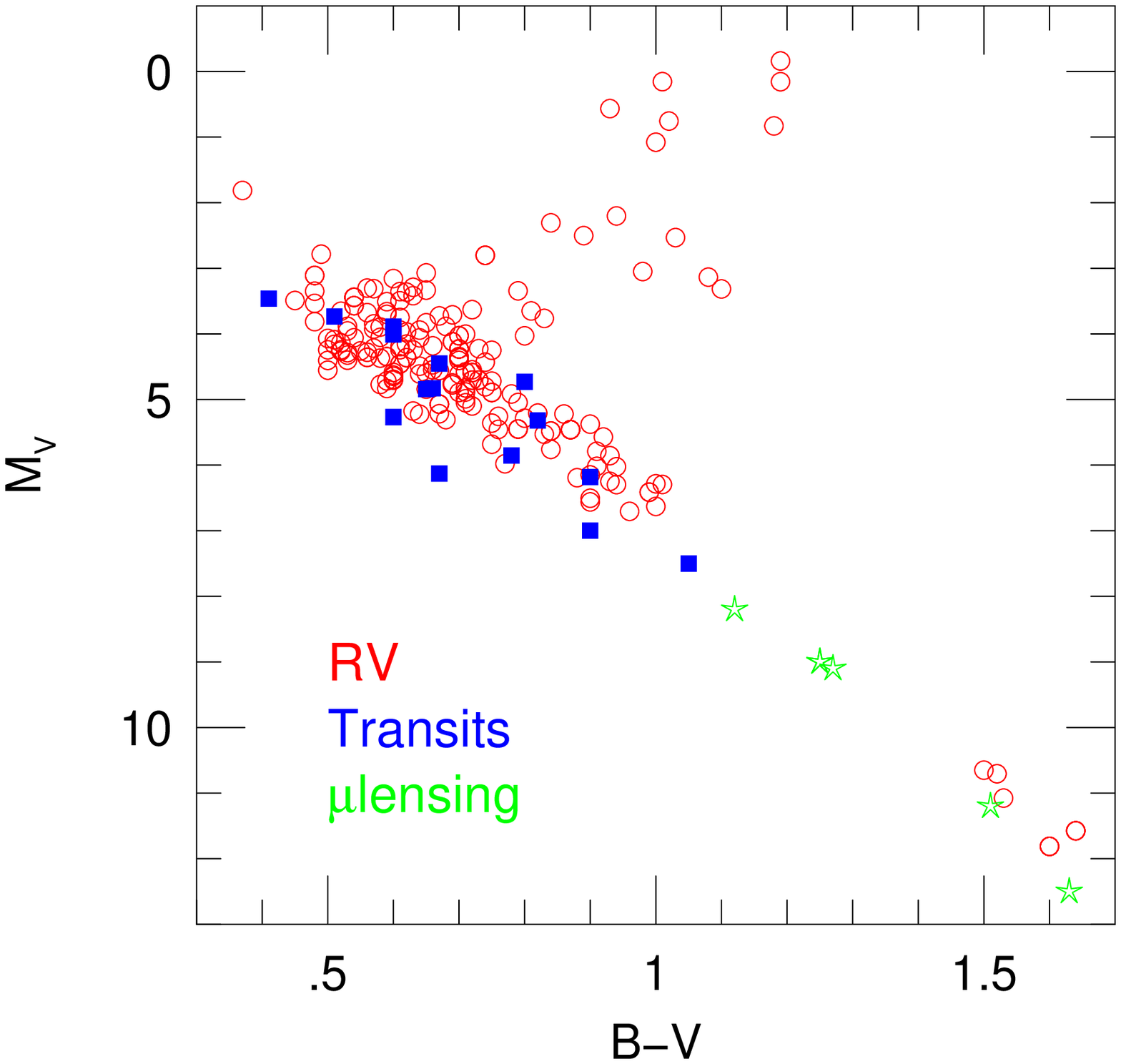}
\caption{\label{fig:cmd}
Color-magnitude diagram of the host stars of planets
detected by the Doppler (RV), transit, and microlensing techniques,
as of June 2007.}
\end{figure}

\begin{figure}
\plotone{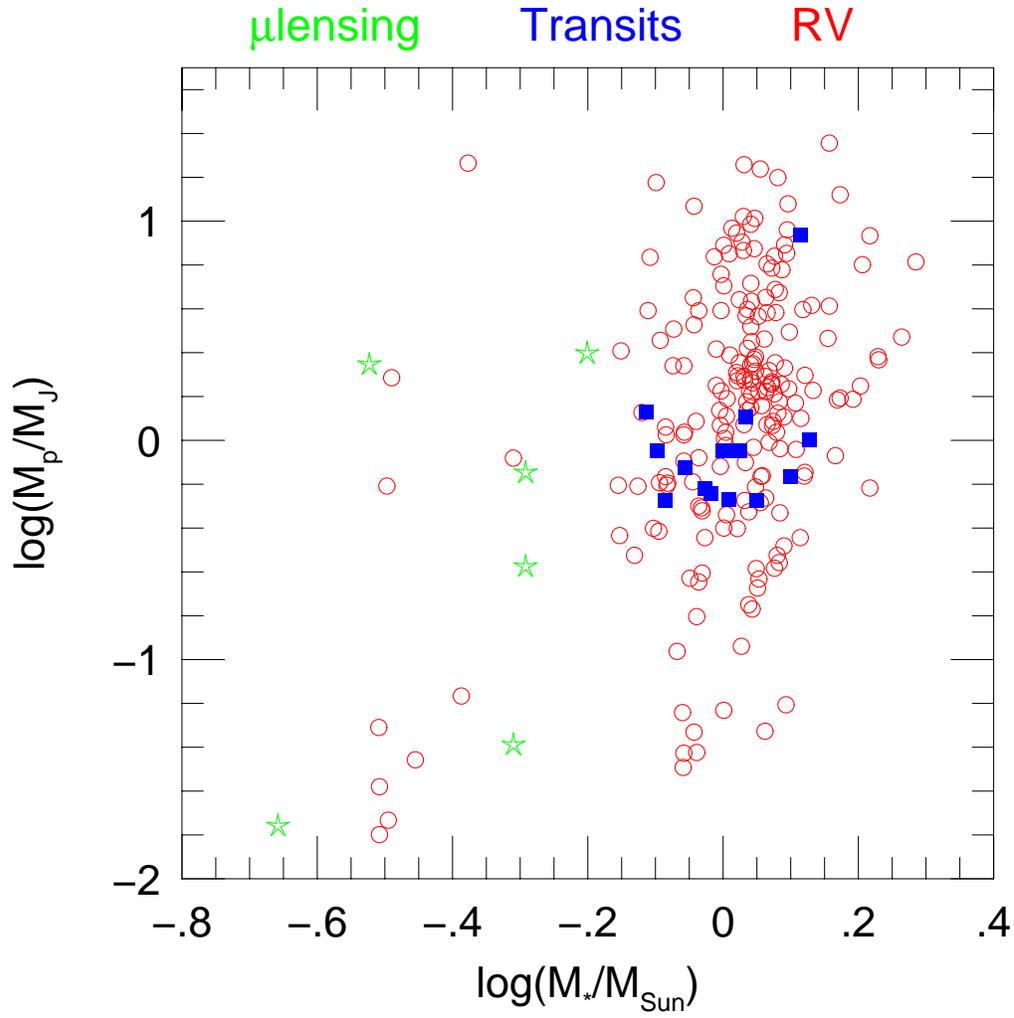}
\caption{\label{fig:mm}
Planet mass vs.\ host mass for the set of targets shown in Fig.\ \ref{fig:cmd}.}
\end{figure}

At yet higher masses, this technique becomes less effective:
declining stability of the red-giant atmospheres and increasing
stellar mass both tend to decrease the size of orbits that
can be probed, while increasing red-giant radii remove close-in
planets.  Finally, the declining mass function and declining
lifetime both tend to reduce the overall population of hosts.
To date, there are no planets detected with host masses
$M>2.5\,M_\odot$, although \citet{lovis07} did find a low-mass
brown dwarf orbiting a star with $M=3.9\,M_\odot$.
Figure \ref{fig:cmd} is a color-magnitude diagram of all hosts detected by
the RV, transits, and microlensing as of June 2007.  Figure \ref{fig:mm}
shows the mass of these planets vs.\ the mass of their host stars.

The transit technique is challenged to detect planets around
massive stars for three reasons: first they are big, second they are
rare, and third (most critically) there is no way to confirm detections.
In principle, the first problem could be overcome by more accurate
photometry, which would probe the smaller fractional transits caused
by large stellar radii.  Lack of targets is a more fundamental
problem, since even very large surveys are detecting
relatively few transiting planets even around much more common
stars.  So, in practice, selection pressure drives this technique
to a relatively narrow range of F and  G stars (see Fig.\ \ref{fig:cmd}).
But the key problem is that if a planet were detected, the same
line-lacunae that plague RV in this mass range would make
confirmation extremely difficult.  Perhaps with prodigious effort,
late M-dwarf ``Jupiter impersonators'' could be rejected, but
the planets could not be reliably distinguished from brown dwarfs,
nor could their masses be measured.

Microlensing is often said to be insensitive to host mass, and indeed
Figure \ref{fig:cmd} shows that microlensing host masses trace the MS
mass function: low mass hosts dominate.  However, this very fact
makes microlensing relatively insensitive to high-mass hosts, just
exactly because of their low frequency of birth and early demise.
Moreover, microlensing's ``insensitivity to host mass'' is predicated
on the fact that all potential GKM hosts are fainter (or at least not
much brighter) than the bulge sources (which are typically
turnoff stars, with some subgiants and giants).  By contrast,
massive stars are substantially brighter than typical microlensed
sources, greatly reducing the probability that the underlying
microlensing event could be detected at all under the ``glare''
\citep{einstein36} of the lens.

Of course, 3 planets have been detected around a pulsar \citep{wolszczan91},
which is a remnant of a massive star, $M>8\,M_\odot$.  However,
it is difficult to see how planets could survive a supernova explosion
without becoming unbound, so these planets are believed to have
formed out of an accretion disk generated by fallback from the explosion.
Thus, while an extremely interesting system, the pulsar planets probably
do not provide any direct information on planets around massive stars.

\section{A New Approach: Massive White Dwarfs}

The very factors that make it difficult to find planets around massive
stars $3\,M_\odot \la M \la 7\,M_\odot$, actually facilitate their detection
when the hosts have evolved into white dwarf (WD) remnants.  
Their short MS lifetimes imply that their planets are still relatively hot (and so luminous)
at the time they become WDs.  The WDs inherit the huge reservoir of
heat of their progenitors, making them both luminous and blue,
both of which facilitate their discovery as ``contaminants'' in
magnitude-limited quasar surveys.
Finally, these relatively massive
WD remnants are physically small ($r_{\rm WD}\sim r_\oplus$, 
\citealt{hamada61}), which
sharply limits their luminosity in the infrared (IR), where the
planet spectrum peaks.  

The range of original periods that can survive the conversion
of a massive star into a white dwarf is roughly
$4\,{\rm yr}< P_{\rm orig}< 10^4\,{\rm yr}$, corresponding
roughly to semi-major axes $3.5\,{\rm AU}< a_{\rm orig}<1000\,{\rm AU}$
\citep{villaver07}.
For smaller orbits, the planet will be swallowed by its host
during its red giant phase.  For larger orbits, the planet
will become unbound because the star can eject half its mass
on timescales of roughly $10^4$ -- $10^5$ years (depending on mass).  
At intervening orbital separations, the planet
will evolve adiabatically to larger radii according to
the formula $a_{\rm fin}/a_{\rm orig} = M/M_{\rm WD}$, still
an order of magnitude smaller than orbits that could be disrupted
by Galactic tides.

\section{Palomar-Green Sample}

To test the feasibility of searching for planets around the WD remnants
of massive stars, we have identified 51 hot, young WDs in the mass range 
$0.7\,M_\odot<M_{\rm WD}<1.0\,M_\odot$, found in the Palomar-Green (PG)
survey.  \citet{liebert05} present a detailed spectroscopic analysis of
these WDs for which they obtained accurate mass and age estimates.

According to the models of \citet{burrows03}, planet luminosity is
heavily suppressed at 3.6\,$\mu$m and 5.8\,$\mu$m, implying that
warm planets will be most easily visible by {\it Spitzer} at 4.5\,$\mu$m.
We therefore further restrict the PG sample to those WDs above
$20\,\mu$Jy at 4.5\,$\mu$m.  This eliminates just 2 targets, leaving
a sample of 49.

\begin{figure}
\plotone{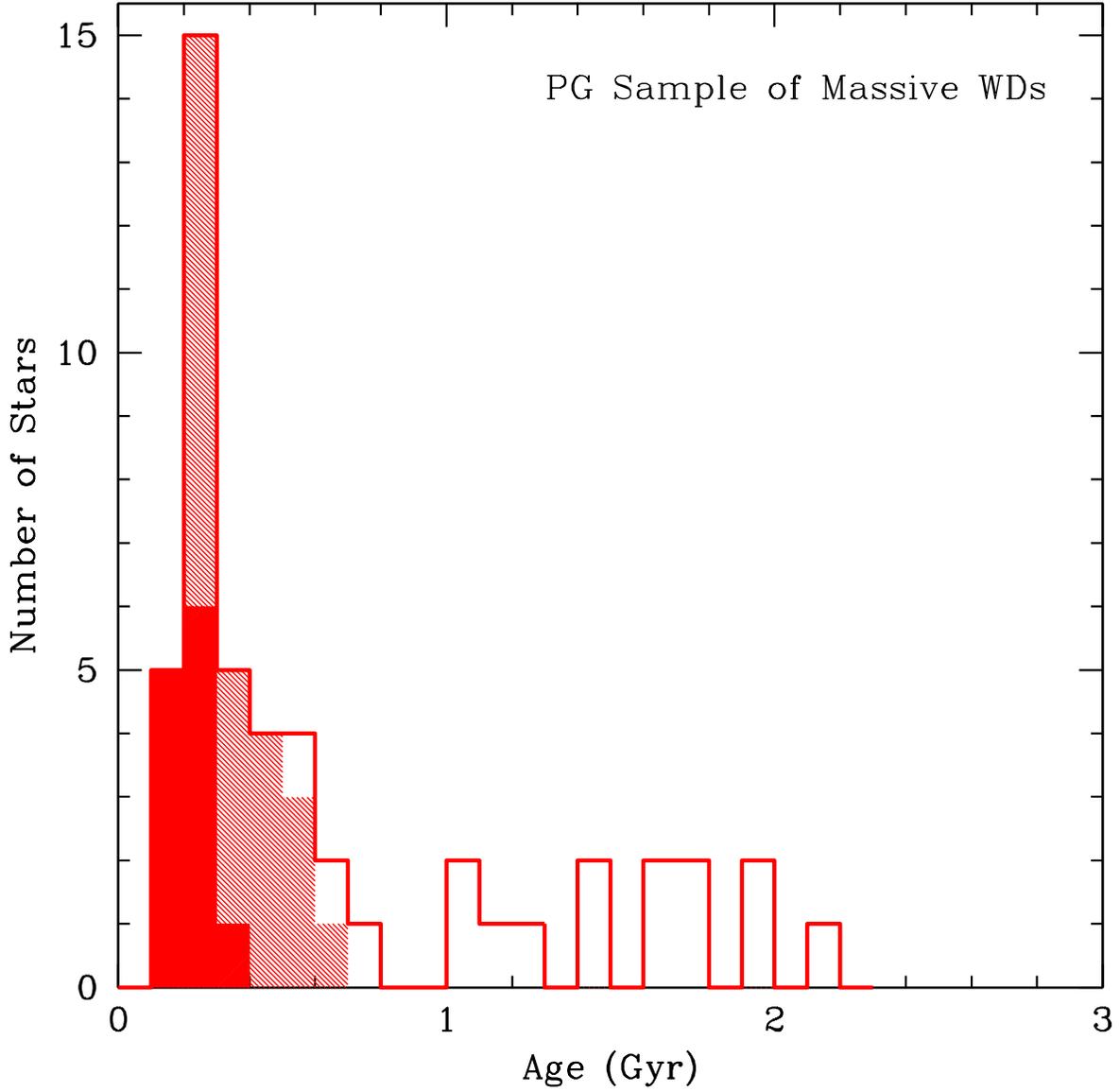}
\caption{\label{fig:ages}
Age distribution of WDs, 
$0.7\,M_\odot < M_{\rm WD} < 1.0\,M_\odot$ from the PG sample.  
The contribution from 30000~K, 20000~K, and 10000~K white dwarfs
are shown as solid, hatched, and empty histograms, respectively.}
\end{figure}

Figure \ref{fig:ages} shows the age distribution of these WDs.
Here ``age'' is defined as time since main-sequence
(and so presumably planet) birth.
Note that the sample is peaked at 300 Myr and that the
great majority are younger than 1 Gyr.

\begin{figure}
\plotone{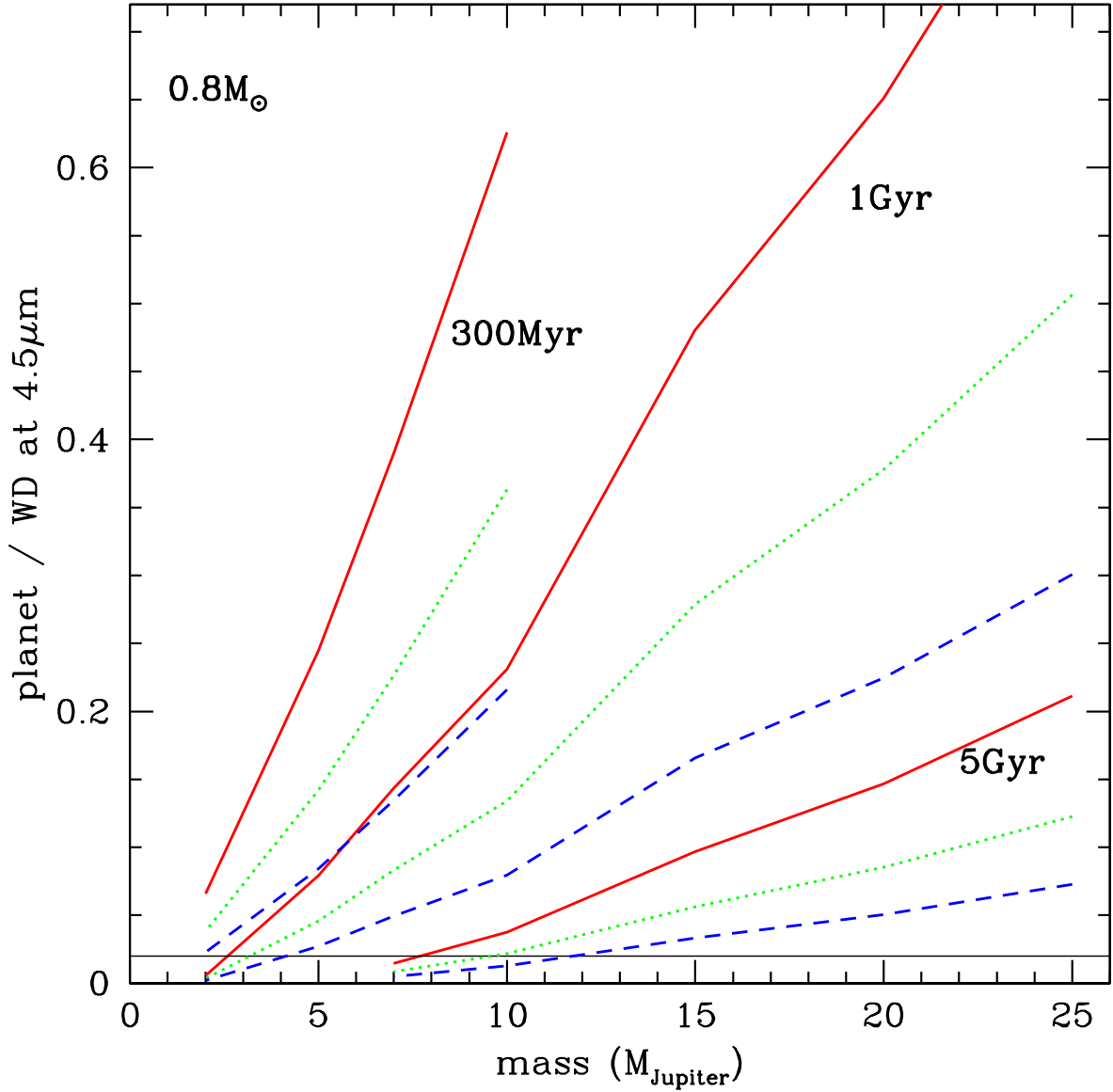}
\caption{\label{fig:detectability} Expected 4.5$\mu$m excess for
planets as a function of mass, at cooling times of 300 Myr, 1 Gyr, and
5 Gyr.  The predicted excess for 10000~K, 20000~K, and 30000~K WDs
are shown as red solid lines, green dotted lines, and blue
dashed lines, respectively.  The horizontal black solid line shows the
2\% absolute calibration uncertainty for the {\it Spitzer} IRAC instrument
\citep{reach05}.}
\end{figure}

Youth is important because it is the young
planets that provide the greatest chance of detection.
Figure \ref{fig:detectability} 
shows planet/WD flux ratio at 4.5 $\mu$m as a function
of planet mass (vertical  axis) and for various ages and WD temperatures,
using \citet{burrows03} models.  A detection requires that the
WD IR flux be accurately predicted based on the temperature and
angular radius derived from flux measurements at shorter wavelengths,
so that the excess flux due to the planet can be inferred from the
{\it Spitzer} measurement. The fundamental limit of the technique
is therefore set by the 2\% error in the
IRAC absolute flux-calibration (horizontal solid line, \citealt{reach05}).

Note that there is potentially good sensitivity to few-Jupiter mass
planets for the majority of the WD ages that are shown in Figure \ref{fig:ages}.
Such planet masses are not uncommon among the red-giant targets of
RV surveys, which have somewhat lower-mass MS progenitors than these
WDs (see Fig.\ \ref{fig:mm}).  \citet{johnson07} find that the
Jovian-planet fraction increases from
1.8\% for $0.1-0.7M_\odot$ stars to 4.2\% for Sun-like stars, to 
8.9\%, for $1.3-1.9M_\odot$ stars for $a<2.5\,$AU and $m>0.8\,M_{\rm jup}$.
Of course,
the analogs of all of these planets around higher mass stars would be
swallowed before they evolved into WDs, and the frequency of planets
around massive stars at wider separations is completely unknown.  However,
if there are comparable numbers at wider separations and if
the observed trend continues or even flattens
at higher masses, then the PG sample would be expected to contain 
several detectable planets: roughly 
$49\times 8.9\%\times(\log(13/2)/\log(13/0.8) = 2.9$.  Here we have
assumed a detection threshold of $m=2\,M_{\rm jup}$ and that planet
masses are distributed as $dN/dm\sim m^{-1}$ from the \citet{johnson07}
threshold to the brown-dwarf limit, $0.8<m/M_{\rm Jup}<13$.

The WDs are quite bright, $V\la 16$, so obtaining the accurate optical/near-IR
photometry required to predict the fluxes in the IRAC bands would be 
straightforward.  Indeed, 41 of the 49 PG WDs already have SDSS photometry 
and 47 are detected in $J$ band by 2MASS.

\section{Planets vs.\ Disks}

Of course, an IR excess at $4.5\,\mu$m could have causes other than planets,
in particular, circumstellar disks. At present, these alternatives
can be easily distinguished by observing the WDs at the other
{\it Spitzer} IRAC bands: warm-planets would only show an excess at $4.5\mu$m, while
cool-dust emission would be detected in most IRAC bands. In fact, all but one of the WDs with debris
disks show excess emission in all 4 IRAC bands (von Hippel et al. 2007; Jura et al. 2007), and the majority of them
also show excess in the K-band (Kilic et al. 2006). There is only one WD, G166-58, known to have a debris disk that
shows up as an excess redward of $5\mu$m (Farihi et al. 2007).
After {\it Spitzer}'s cryogen is exhausted, it will still be possible to differentiate planets
from {\it warm} debris disks since these disks would show excesses both in the $3.6\mu$m and $4.5\mu$m bands.
However, the channel for discriminating between planets and cooler disks will disappear.

The candidates found by ``Warm {\it Spitzer}'' from observations at $3.6\,\mu$m and $4.5\,\mu$m can still be distinguished from cool disks
in some cases. The median distance of the PG WDs is 68 pc,
implying that planets lying at projected separations
$r_\perp\ga 140\,\au$
would be separately resolved by {\it Spitzer}.  Such planets would 
be recognized by their lack of optical counterpart (in SDSS for the
regions covered by that survey). Because the planet orbit
expands by $a_{\rm final}/a_{\rm orig}=M/M_{\rm WD}$, i.e., a factor 5--7,
these separations correspond to $a_{\rm orig}\ga 20-25\,\au$,
which is larger than the orbits of Jovian planets in the Solar System.
Moreover, there is a correlation between the ages and distances of the targets. Since the PG survey is magnitude-limited,
hotter (younger) objects can be detected at larger distances. The median distances for PG WDs with ages $\leq300$ Myr
and $\leq1$ Gyr are 106 pc and 89 pc, respectively. Hence, planets at orbital separations $a_{\rm orig}\ga 35-40\,\au$ would be resolved
with {\it Spitzer} around the WDs younger than 1 Gyr.
\citet{kasper07} failed to find any 
planets of $m>2\,M_{\rm jup}$ and
$a>30\,\au$ around 22 young GKM stars, while \citet{lafreniere07} failed
to find any at $a>40\,\au$ around 85 young GKM stars.  However,
the orbits of planets around massive stars could be larger.

The {\it James Webb Space Telescope (JWST)} could in principle confirm
most of the WD planets detected by {\it Spitzer} observations.
The {\it JWST} Near-Infrared Camera (NIRCAM) FWHM at $4.5\,\mu$m is $\sim0.15''$.  Since the planet/WD
flux ratios at this wavelength must be greater than 2\%, resolution
at 1 FWHM is feasible. At the median sample distance ($D=68$ pc), the
minimum allowed orbit $a_{\rm final}>10\,\au$ (or $a_{\rm orig}\ga 2\,\au$) corresponds to a slightly
bigger angle. Of course, half the WD sample is at greater distances and,
depending on orbital phase and inclination,
some planets will be significantly closer than $a/D$, but the planets
will not necessarily cluster at the closest allowed semi-major axis
and brighter planets will be detectable at somewhat closer separations
or from the astrometric offset of the combined light as seen in the
optical and IR.  For example, for flux ratios of 10\% and separations
of $0.1''$, the offset is 10 mas, which is easily detectable.

In the meantime, confirmation would be possible for the younger, 
more massive planets using
$H$ band adaptive optics observatons on large ground-based telescopes. 
For example, at 300 Myr, a $m=5\,M_{\rm jup}$ planet has $M_H=19.3$
\citep{burrows03}, compared to $M_H=11.8$ for its WD host, i.e. a flux
ratio of 1000.  Even without coronographs, such ratios should be
resolvable at 5 FWHM, corresponding to $0.3''$ (Van Dam et al. 2006). However, at 1 Gyr,
the same planet is 10 times fainter, while at 300 Myr, a $m=2\,M_{\rm jup}$
is almost 100 times fainter.  Hence, this approach to
confirmation would be extremely
challenging for the majority of detectable planets.

\section{Targets in Other Directions}

\begin{figure}
\includegraphics[angle=0,scale=.7]{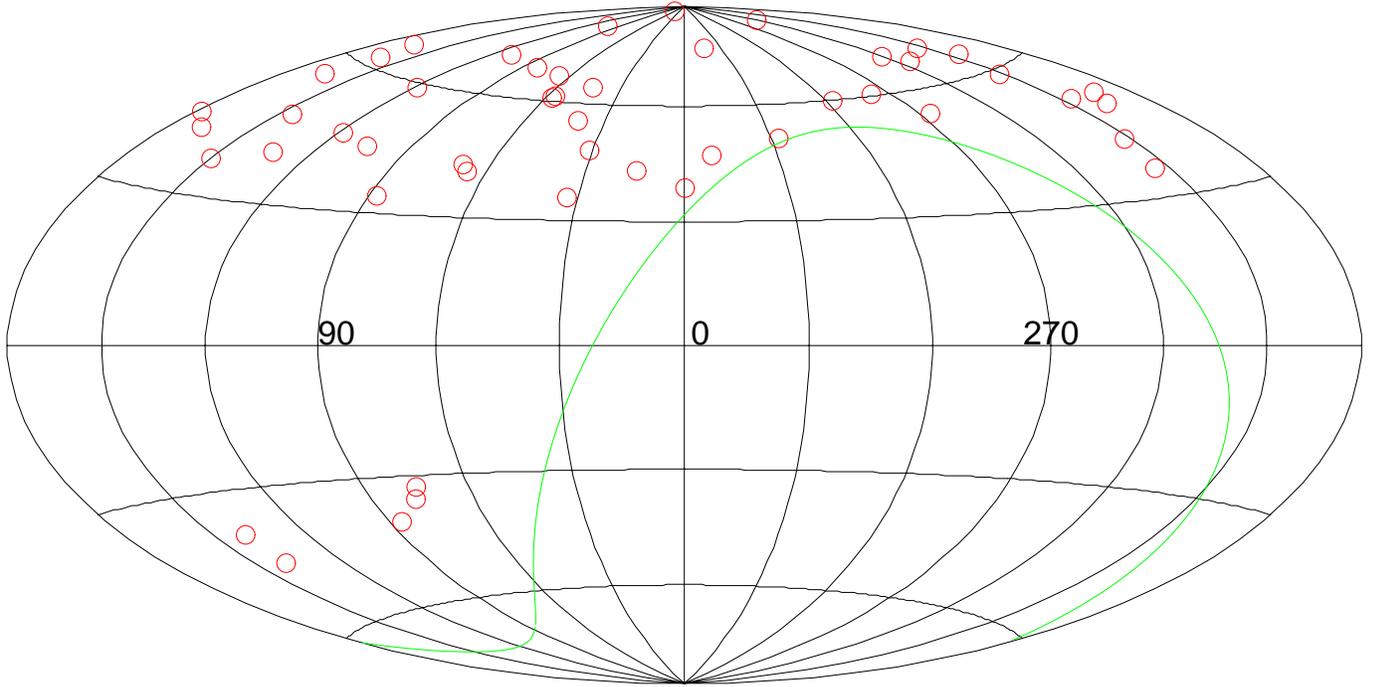}
\caption{\label{fig:aitoff} Distribution of 49 PG massive WDs in Galactic
coordinates.  The green curve is at $\delta=-10^\circ$, the apparent
southern boundary of the survey.  Evidently, about 1/4 of the sky has
been covered.  There is a slight tendency toward lower density near
the Galactic pole, implying that the fields $|b|<30$ will be somewhat
richer in massive WDs than the PG survey area.}
\end{figure}

Figure \ref{fig:aitoff} shows the distribution of the PG sample in
Galactic coordinates.  The green curve is the locus of $\delta=-10^\circ$,
the apparent southern boundary of the survey. Evidently, about 1/4 
of the sky has been covered. There appears to be a slight 
tendency toward a lower density of massive WDs near the Galactic pole.
Such an under-density is expected.  The median distance of the WDs
is $\sim 68\,$pc.  If this distance were small compared to the scaleheight
of the massive WD population, then they should be found isotropically.
However, with typical ages (since MS birth) of about 300 Myr, massive
WDs should be distributed similarly to A stars, which have a scaleheight
of 90 pc \citep{miller79}.  Since this is comparable to the sample distance,
we expect some suppression at the poles.  This would 
imply that the fields $|b|<30$ 
will be somewhat richer in massive WDs than the PG survey area.
Sixty eight per cent of the massive white dwarfs detected in EUVE and ROSAT all-sky surveys are within
$\pm30^{\circ}$ Galactic latitude.
\citet{liebert05} pointed out that the enhanced density of massive white dwarfs at low Galactic
latitudes might be due to some of them being the progeny of B stars in Gould's Belt.

Of course, it is also possible to survey for fainter WDs, but planet
detection will generally be more difficult for these because the WDs
are more distant, so their planets will be both fainter and at smaller
angular separations from the hosts.

\section{Discussion}

Probing WDs for sub-stellar companions began with brown dwarf searches (Probst 1980; Zuckerman \& Becklin 1987; Farihi et al. 2005).
\citet{winget03} proposed to search for planetary companions from timing residuals
of pulsating white dwarfs. \citet{mullally07a} reported a possible detection of a planet around a
pulsating white dwarf, however follow up observations are required to confirm that the observed
change in the pulsation periods is not due to the cooling of the star itself.
\citet{burleigh02,burleigh06}, \citet{debes05,debes07}, Livio et al. (2005), Friedrich et al. (2006), and \citet{mullally07b}
proposed and looked for planetary companions from
excess near- and mid-IR emission. \citet{hansen06} searched for excess IR
emission from disks of very massive WDs, $M_{\rm WD}>1\,M_\odot$.

What is specifically new to this paper is the idea of probing
the previously unexplored stellar mass regime 
$3\,M_\odot \la M \la 7\,M_\odot$
by looking for the planets around the WD remnants of these stars.
Typical field WDs have masses $M_{\rm WD}\sim 0.6\,M_\odot$ and so
have MS progenitors $M\sim 2\,M_\odot$.  This is a very interesting
mass range, currently probed only from red-giant RV studies,
which (as noted above) are restricted in sensitivity to relatively
close companions $P\la 2\,$yr.  Hence, thermal-excess searches
could potentially probe new regimes of parameter space for these
stars.  The very massive WDs studied by \citet{hansen06} are believed
to be the result of WD-WD mergers.  If planets survived this merger
process, it would be both surprising and very interesting.
Of course, some of the massive white dwarfs in the PG sample may be the products
of WD-WD mergers as well. These mergers would result in rapid rotation rates, which could be
measured from the non-LTE cores of the Balmer lines. Massive WDs with planetary companions could
be spectroscopically checked for high rotation rates in order to see if they are the products of binary mergers or single star evolution.

The regime we have targeted has not been systematically
explored. The \citet{mullally07b} sample does contain about a dozen WDs in
this mass range, but only a few of these have ages less than 1 Gyr,
when planets of a few Jupiter masses can still be robustly detected.
We have identified a sample of 49 massive WD targets, the great majority 
of them young, and suggested that 3--4 times as many could be found
in other areas of the sky.
 
There is no guarantee that the observed correlation between stellar 
mass and planet occurrence will continue to these higher masses.
In fact, \citet{kennedy07} suggest that the frequency function
declines for $M>3\,M_\odot$.  Only observations can settle this
question.

\acknowledgments
We thank Jay Farihi, Guillermo Gonzalez,
Mike Jura, Christophe Lovis, Jean Schneider, 
Eva Villaver, Ted von Hippel, Hans Zinnecker, and an anonymous referee for making many
useful suggestions and comments.
This work was supported by NSF grant AST 042758.

\clearpage

\begin{thebibliography}{99}

\bibitem[Burleigh et al.(2002)]{burleigh02} Burleigh, M.~R., Clarke, F.~J., \& Hodgkin, S.~T.\ 2002, \mnras, 331, L41 

\bibitem[Burleigh et al.(2006)]{burleigh06} Burleigh, M., Hogan, E., \& Clarke, F.\ 2006, The Scientific Requirements for Extremely Large Telescopes, 232, 344 

\bibitem[Burrows et al.(2003)]{burrows03} Burrows, A., Sudarsky, D. 
\& Lunine, J.I. 2003, \apj, 596, 587

\bibitem[Debes et al.(2005)]{debes05} Debes, J.~H., Sigurdsson, S., \& Woodgate, B.~E.\ 2005, \apj, 633, 1168 

\bibitem[Debes et al.(2007)]{debes07} Debes, J.~H., Sigurdsson, S., \& Hansen, B.\ 2007, \aj, 134, 1662 

\bibitem[Einstein(1936)]{einstein36} Einstein, A. 1936, Science, 84, 506

\bibitem[Farihi et al.(2005)]{2005ApJS..161..394F} Farihi, J., Becklin, E.~E., \& Zuckerman, B.\ 2005, \apjs, 161, 394 

\bibitem[Farihi et al.(2007)]{2007arXiv0710.0907F} Farihi, J., Zuckerman, 
B., \& Becklin, E.~E.\ 2007, ApJ, in press

\bibitem[Fischer \& Valenti(2005)]{fischer05} Fischer, D.A. \& Valenti, J.
\apj, 622, 1102

\bibitem[Friedrich et al.(2006)]{fri06} Friedrich, S., Zinnecker, H., Correia, S., Brandner, W., Burleigh, M., \& McCaughrean, M. 2006,
ASP Conference Series, 999

\bibitem[Frink et al.(2002)]{frink02} Frink, S., Mitchell, D.S., 
Quirrenbach, A., Fischer, D.A., Marcy, G.W., \& Butler, R.P.\ 2002, 
\apj, 576, 478


\bibitem[Hamada \& Salpeter(1961)]{hamada61} Hamada, T. \& Salpeter, E.E.\ 
1961, \apj, 134, 683

\bibitem[Hansen et al.(2006)]{hansen06} Hansen, B.M.S., Kulkarni, S. \& 
Wiktorowicz, S. 2006, \aj, 131, 1106

\bibitem[Johnson et al.(2007)]{johnson07} Johnson, J.A.; Butler, R.P., 
Marcy, G.W., Fischer, D.A., Vogt, S.S., Wright, J.T., \& Peek, K.M.G.,
2007, \apj, 670, 833

\bibitem[Jura et al.(2007)]{2007ApJ...663.1285J} Jura, M., Farihi, J., \& 
Zuckerman, B.\ 2007, \apj, 663, 1285 

\bibitem[Kasper et al.(2007)]{kasper07} Kasper, M., Apai, D., Janson, M., \& 
Brandner, W. 2007 \aap, 472, 321

\bibitem[Kennedy \& Kenyon(2007)]{kennedy07}Kennedy, G.M., \& Kenyon, S.J.\ 
2007, \apj, in press (arXiv:0710.1065)

\bibitem[Kilic et al.(2006)]{2006ApJ...646..474K} Kilic, M., von Hippel, 
T., Leggett, S.~K., \& Winget, D.~E.\ 2006, \apj, 646, 474 

\bibitem[Lafreni\`ere et al.(2007)]{lafreniere07} Lafreni\`ere, D. et al. 2007,
\apj, 670, 1367

\bibitem[Laws et al.(2003)]{laws03}Laws, C., Gonzalez, G., Walker, K.M., 
Tyagi, S., Dodsworth, J., Snider, K., \& Suntzeff, N.B.
2003, \aj, 125, 2664

\bibitem[Liebert et al.(2005)]{liebert05} Liebert, J., Bergeron, P. 
\& Holberg, J.B. 2005, \apjs, 156, 47 

\bibitem[Livio et al.(2005)]{2005ApJ...632L..37L} Livio, M., Pringle, 
J.~E., \& Wood, K.\ 2005, \apjl, 632, L37 

\bibitem[Lovis \& Mayor(2007)]{lovis07} Lovis, C. \& Mayor, M. 2007, 
\aap, 472, 657

\bibitem[Miller \& Scalo(1979)]{miller79}
Miller, G.E. \& Scalo, J.M.\ 1979, \apjs, 41, 513

\bibitem[Mullally et al.(2007a)]{mullally07a} Mullally, F., Winget, D.~E., \& Kepler, ~S.~O.\ 2007, ASP Conference Series, 372, 363 

\bibitem[Mullally et al.(2007b)]{mullally07b} Mullally, F., Kilic, 
M., Reach, W.~T., Kuchner, M.~J., von Hippel, T., Burrows, A., \& Winget, 
D.~E.\ 2007, \apjs, 171, 206 

\bibitem[Probst(1980)]{1980BAAS...12..507P} Probst, R.~G.\ 1980, \baas, 12, 507 

\bibitem[Reach et al.(2005)]{reach05} Reach, W.T., et al.\ 2005, \pasp, 117, 978

\bibitem[Reffert et al.(2006)]{reffert06}Reffert, S., Quirrenbach, A., 
Mitchell, D.S., Albrecht, S., Hekker, S., Fischer, D.A., Marcy, G.W., \& 
Butler, R.P.\ 2006, \apj, 652, 661

\bibitem[van Dam et al.(2006)]{2006PASP..118..310V} van Dam, M.~A., et al.\ 2006, \pasp, 118, 310

\bibitem[Villaver \& Livio(2007)]{villaver07} Villaver, E. \& Livio, M. 2007,
\apj, 661, 1192

\bibitem[von Hippel et al.(2007)]{2007ApJ...662..544V} von Hippel, T., Kuchner, M.~J., Kilic, M., Mullally, F., \& Reach, W.~T.\ 2007, \apj, 662, 544 

\bibitem[Winget et al.(2003)]{winget03} Winget, D.~E., et al.\ 2003, Scientific Frontiers in Research on Extrasolar Planets, 294, 59 

\bibitem[Wolszczan(1991)]{wolszczan91} Wolszczan, A.\ 1991, BAAS, 23, 1347

\bibitem[Zuckerman \& Becklin(1987)]{1987ApJ...319L..99Z} Zuckerman, B., \& 
Becklin, E.~E.\ 1987, \apjl, 319, L99 

\end{thebibliography}
\end{document}